\documentclass[aps,pre,twocolumn,amsmath,amssymb]{revtex4}
\usepackage{bm}
\usepackage{graphicx}

\newcommand{\be}{\begin{eqnarray}}
\newcommand{\ee}{\end{eqnarray}}

\newcommand{\ba}{\begin{array}}
\newcommand{\ea}{\end{array}}

\newcommand{\ket}[1]{|#1\rangle}

\begin{document}

\title{Monogamy of entanglement as a necessary and sufficient condition for safe QKD in any physical theory}
\author{Marcin Paw{\l}owski}
\address{ Institute of Theoretical Physics and Astrophysics, Uniwersytet Gda\'{n}ski, PL-80-952, Gda\'{n}sk
\\
Katedra Fizyki Teoretycznej i Informatyki Kwantowej, Politechnika Gda\'{n}ska, PL-80-952, Gda\'{n}sk}

\begin{abstract}
We show that the monogamy of entanglement is a sufficient phenomenon in every physical theory, if the quantum
key distribution is to be safe on the grounds of such theory. To do so we present the QKD protocol that is safe
in any physical theory under the assumption of the monogamous entanglement only. The necessity of this condition
is also discussed.
\end{abstract}

\maketitle

\section{Introduction}

The quantum cryptography offers us unprecedented degree of safety, with the laws of quantum mechanics serving as a
warranty. Recently the issue of what happens to this safety beyond quantum mechanics has been raised \cite{BHK,Nasza}.
Both these papers present protocols that are safe in the classes of theories that include but are not confined to
quantum mechanics and both have their disadvantages. In \cite{BHK} authors base the security of their protocol on the
assumption of no signalling while repeating after \cite{Val2} that every deterministic non-local theory allows signalling.
This excludes the wide range of the most promising alternatives to the quantum mechanics, that is the non-local,
hidden variables theories. The protocol presented in \cite{Nasza} on the other hand is proven secure even in the light
of de'Broglie-Bohm theory with the initial particle positions known to Eve. Unfortunately it is the only non-local, realistic
theory for which the proof in \cite{Nasza} can be validated. The disadvantage shared by both these protocols is that
their implementation would be very difficult. In this paper we present the entangled-state
cryptographic protocol that is not only secure in any non-local, hidden variables theory which involves the monogamy
of entanglement but also much easier to implement that those discussed above.
The area of this protocol's safety includes but is not confined to the classes of theories covered by
the protocols of \cite{BHK,Nasza}.
In fact it is shown that this protocol is safe in every physical theory where safe QKD is at all possible.

\section{Limitations}
We start by setting the necessary limitations on communicating parties and eavesdropper. Though Eve can have the knowledge and technology
to exploit some yet unknown post-quantum theory, Alice and Bob are limited to the use of quantum mechanics at its present form.
The theory used by the eavesdropper must also have three important traits in order to allow for safe QKD of any kind. Firstly, Eve can
not have any influence on or knowledge of the choices of the measurements being made at Alice's and Bob's labs. Secondly, the theory
exploited by Eve must agree with QM on the outcomes of all the experiments so far conducted. Finally, the entanglement must be monogamous.
The necessity of this condition is quite obvious. If the
entanglement was not monogamous there would be no way for Alice and Bob to ascertain whether their systems are entangled with
some external party or not. That party could made the measurements after the bases are revealed and get any information
that is needed. For the proof of sufficiency we present QKD protocol which safety is based only on the limitations mentioned in this section.
\

Before we start we should clarify what is meant by the monogamy of entanglement. We will assume that the monogamy of entanglement is a
property of a theory that disallows a third party to find anything about the outcomes of the measurements done on a maximally entangled state
by measuring some other system (other ways of finding the outcomes are not forbidden). By a maximally entangled state we will mean a state
for which two parties can always expect perfect correlations or anti-correlations even if they can choose between the measurements in two 
mutually unbiased bases (as long as both parties choose the same base).

\section{Protocol}
Consider the following protocol:

\

1. Alice and Bobs receive their parts of the system in a singlet state.
They both randomly choose two parameters $c=0,\frac{\pi}{2}$ and $\phi=0,\frac{\pi}{8},\frac{\pi}{4},\frac{3\pi}{8},\frac{\pi}{2}$
and measure their parts of the system in basis
\be
\ket{+1}=\cos(\phi+c)\ket{0}+\sin(\phi+c)\ket{1}
\\
\ket{-1}=\cos(\phi+c)\ket{1}-\sin(\phi+c)\ket{0}
\ee
They repeat this whole step for all the states that they receive.

\

2. They randomly choose an appropriate part of the runs of the protocol.
In this cases they both reveal their bases and measurement
outcomes and calculate the expectation value from the CHSH inequality (basing on the runs where the choice of their bases was right for
this inequality). They abort the protocol if it differs from the predictions of quantum mechanics
for the singlet state, that is $2\sqrt{2}$. They also abort the protocol if they do not observe perfect anti-correlations for the same
choice of the bases and perfect correlations for the choice of bases differing by $\frac{\pi}{2}$.

\

3. They announce their choices of $\phi$ (but not $c$) for each of the remaining cases.

\

4. For each of the remaining cases they randomly choose one of the parties.
This party reveals its $c$ and a random bit XORed
with the measurement outcome (of course binary numbers have to be first assigned to the outcomes).
Second party has now the knowledge of both bases so he/she knows whether the results are
correlated or anti-correlated and knowing his/her own result can decode the random bit send.

\

5. Step 4 is repeated $K$ times, so both parties share a set of $K$ bits. From this set one bit of key is generated by
XORing all the bits form this set.

\section{Proof of safety}

If we assume that the standard quantum mechanics is right then the Step 2 is enough
to be sure that Eve has no knowledge of the state
Alice and Bob receive and {\it ipso facto} the key. The protocol is then secure as proved in \cite{BBM}, where the perfect correlations in two
mutually  unbiased bases are exploited to prove that the state received is indeed maximally entangled.
Another way to prove the security on the grounds of quantum mechanics is to notice that Step 2
uses similar measures to the Ekert's original paper \cite{E91}.
The proof of security for any nonlocal hidden variable
model that includes the monogamy of the entanglement is more difficult. We assume that the theory known to
the eavesdropper is deterministic since every probabilistic theory can be simulated by an appropriate
deterministic model.
\

Since we assume the monogamy of the entanglement,
the only way for Eve to know the results of the experiment is to have the access to some hidden
initial parameters and know the mechanics (non-local in principle) that govern the evolution of the state,
so she knows what the results of Alice's and Bob's measurements will be for any choice of the bases.
The only thing that she does not know is what bases were actually
chosen, since at each time only one of them is revealed.
This type of attacks has already been presented in \cite{Val} and \cite{Plaga} (actually these are the only two papers the author is
aware of that propose attacks that are beyond the scope of current physical theories). The hidden variables in this papers
are particle positions and gravitational fields respectively. Since \cite{Plaga} deals with the BB84 \cite{BB84}
protocol, the method presented there does not apply to the safety of the protocol above. Even if Plaga's method could be
generalized, the QKD protocol presented here would remain secure against the type of attacks presented in \cite{Plaga}
as well as in \cite{Val}.

Let us now assign the numbers $\pm1$ to the experiment outcomes. Eve knows the full structure of the functions
$W_A(\lambda,a,b)$ and $W_B(\lambda,a,b)$ which govern the results at Alice's and Bob's labs. She also knows hidden
parameter $\lambda$ in each case, $a$ and $b$ denote the bases of Alice and Bob respectively.
Since perfect (anti-) correlations are expected in each case (Step 2 guarantees that) these functions must obey
\be
W_A(\lambda,a,0)W_B(\lambda,a,0)=-W_A(\lambda,a,\frac{\pi}{2})W_B(\lambda,a,\frac{\pi}{2})
\\
W_A(\lambda,0,b)W_B(\lambda,0,b)=-W_A(\lambda,\frac{\pi}{2},b)W_B(\lambda,\frac{\pi}{2},b)
\ee
which is equivalent to
\be \label{pl}
W_A(\lambda,a,0)W_A(\lambda,a,\frac{\pi}{2})=-W_B(\lambda,a,0)W_B(\lambda,a,\frac{\pi}{2})
\\ \label{dl}
W_A(\lambda,0,b)W_A(\lambda,\frac{\pi}{2},b)=-W_B(\lambda,0,b)W_B(\lambda,\frac{\pi}{2},b)
\ee
It is clear form (\ref{pl}) that for every $\lambda$ and $a$ either $W_A$ or $W_B$ changes sign as $b$ changes.
From (\ref{dl}) we see that the same is true for $a$. We can now divide the set of $\lambda$s into two separate
subsets. For $\lambda \in \Lambda_1$,  $W_A(\lambda,a,b)$  does not depend on $b$ and
$W_B(\lambda,a,b)$  does not depend on $a$. We will call this set local. The rest of $\lambda$s are in the
set $\Lambda_2$ which we will call non-local.
\

Let us assume that the parties are now during Step 4 of the protocol and Eve's $\lambda$ corresponding to this system is
form the set $\Lambda_2$. Depending on the chosen party the knowledge of either $W_A$ or $W_B$ is necessary for the decoding
of the random bit. At least one of the outcomes changes with the change of the other party's base that is unknown to
Eve, which means that in every case Eve does not know at least one of the outcomes. If it is the outcome of the chosen party
then Eve cannot decode the random bit. It happens in 50\% of the cases. So if $\lambda \in \Lambda_2$ then Eve has the
chance $0.5$ of knowing the random bit. Now it is necessary to estimate the minimal size of the subset $\Lambda_2$ compared
to the whole set of $\lambda$s.
\

To do that let us notice that the theory is required to give the outcome $2\sqrt{2}$ in Step 2. If all the $\lambda$s were
from the local set this value would be 2. If they all were from the non-local set it could be as high as 4. Now
it is easy to calculate that $\sqrt{2}-1$ of $\lambda$s must be in the non-local set (throughout the whole paper we assume
that the probability for each $\lambda$ to occur is the same - we can do this without any loss of generality).
\

Now we can conclude that Eve can know the random bit with the probability $p=\frac{3-\sqrt{2}}{2}\approx0.79$
(She knows the bit if $\lambda$ is in the local set and in the half of the cases from the non-local one). Since Step 4
is repeated $K$ times then the bit of key generated in Step 5 is known by Eve with the probability $P=p^K$. Increasing $K$
leads to arbitrary small probabilities (for $K=20$, $P<1\%$ and increasing $K$ by 10 reduces P over 10 times).

\section{Discussion}
The protocol that has been presented uses only the assumption of monogamous entanglement and thus proves that it is the
sufficient condition for the safety of quantum key distribution.
\

The necessity and sufficiency of the only assumption used by the protocol means that
the protocol presented is safe in every physical theory  where it is possible to construct one.
Although there is some room for
improvement in the fields of speed and efficiency of the protocol and the safety when
it comes to physical implementations, the safety of this protocol in the terms of physicals principles that guarantee
it is the highest attainable.
\

This work is part of EU 6FP programme QAP. The author thanks M. Bourennane, W. Laskowski,  T. Paterek and M. \.{Z}ukowski for their important input.

\end{document}